\documentclass[11pt]{article} 

\usepackage[utf8]{inputenc} 

\usepackage{geometry} 
\usepackage{enumerate}
\usepackage{amsmath}
\usepackage{amssymb}
\usepackage{dsfont}
\usepackage{amsfonts}
\usepackage{graphicx}
\usepackage{epstopdf}
\usepackage{algorithmic}
\usepackage{verbatim}
\ifpdf
  \DeclareGraphicsExtensions{.eps,.pdf,.png,.jpg}
\else
  \DeclareGraphicsExtensions{.eps}
\fi

\numberwithin{equation}{section}
\numberwithin{figure}{section}
\numberwithin{table}{section}

\allowdisplaybreaks

\title{Extensions of Dupire Formula: Stochastic Interest Rates and Stochastic
Local Volatility}

\author{\Large Orcan \"Ogetbil\thanks{orcan.ogetbil@wellsfargo.com}, \ 
and Bernhard Hientzsch\thanks{bernhard.hientzsch@wellsfargo.com}\\
{\small Corporate Model Risk, Wells Fargo Bank}}

\date{} 

\usepackage{amsopn}

\begin{document}

\maketitle
\setcounter{secnumdepth}{3}

\begin{abstract}
We derive generalizations of Dupire formula to the cases of general stochastic
drift and/or stochastic local volatility. First, we handle a case in which the
drift is given as difference of two stochastic short rates. Such a setting is
natural in foreign exchange context where the short rates correspond to the
short rates of the two currencies, equity single-currency context with
stochastic dividend yield, or commodity context with stochastic convenience
yield. We present the formula both in a call surface formulation as well as
total implied variance formulation where the latter avoids calendar spread
arbitrage by construction. We provide derivations for the case where both
short rates are given as single factor processes and present the limits for a
single stochastic rate or all deterministic short rates. The limits agree
with published results. Then we derive a formulation that allows a more general
stochastic drift and diffusion including one or more stochastic local volatility
terms. In the general setting, our derivation allows the computation and
calibration of the leverage function for stochastic local volatility models.
Despite being implicit, the generalized Dupire formulae can be used numerically
in a fixed-point iterative scheme.
\end{abstract}


\paragraph{AMS}
  91G20, 91G30

\section{Introduction}

Risk neutral pricing frameworks aim to establish methodologies for producing
prices consistent with market data available as of valuation time. As a standard
approach, practitioners consider parametric models to map market quotes to
time and space dependent model parameters. The single parameter
Black-Scholes model, for instance, gives European vanilla option prices
as a function of \emph{implied volatility}. In a sense, having an implied
volatility surface spanning a range of strikes and maturities is equivalent to
knowing the prices of European vanilla options, whose payoffs depend on the
value of the underlier solely at maturity, for the same strikes and maturities.
This in turn amounts to knowing the risk-neutral probability density of the
underlier at given future times conditioned on its present value. In this
paradigm, the bulk of the work in developing a methodology to price European
vanilla option instruments written on the same underlier lies in the
construction of the implied volatility surface.

The benefits of formulating the risk-neutral probability density as a function
of time and underlying spot value, however, go beyond the ability to price
European vanilla options. To price more complex options, whose payoffs depend
not only on the terminal value of the underlier, but also on its intermediate
values, one can make use of the risk neutral densities implied by the market
prices of European vanilla options at the intermediate times.
Dupire \cite{Dupire1994}, and Derman and Kani
\cite{DermanKani1994} showed that there is a unique diffusion process that
implies risk neutral probabilities consistent with the European vanilla option
market quotes. Dupire's formula provides a map in a non-parametric way between
European vanilla option market prices and the diffusion coefficient under the
assumption of deterministic interest rates.

Various authors considered extensions of the local volatility formulation, or
embedding local volatility into more complex hybrid models. These efforts
include incorporating jump-diffusion
processes \cite{AndersenAndreasen2000}, representing the effect of one
stochastic interest rate as a bias to the fully deterministic rate model
\cite{Benhamou2008}, local volatility with single interest rate following
a Vasicek model \cite{Hu2015}, introducing stochasticity to local
volatility \cite{Jex1999, Lipton2002, Tian2012, HMR2014}, and embedding local
volatility into a nonlinear McKean SDE \cite{GH2012}. We refer to these papers
for historical background on the development of local volatility based models.

The primary goal of this paper is to provide a self-contained, detailed
derivation for the case of a drift given as the difference of two short rates
driven by single factor processes, and to provide a new generalization to a case
with very general stochastic drift and diffusion terms. In Section
\ref{sec:simple_model}, we construct the direct extension of the standard local
volatility model to cover stochastic domestic and foreign interest rates in a
foreign exchange (FX) derivatives setting. The result is presented in both
vanilla call option price (\ref{eqn:extendeddupire}) and total Black-Scholes
implied variance (\ref{eqn:extendeddupire_tiv}) formulations. We also consider
the limiting cases, with one or both rates being deterministic, to recover
results that can be found in literature ((\ref{eqn:dupire_singlecall}),
(\ref{eqn:dupire}), (\ref{eqn:dupire_single_tiv}), and
(\ref{eqn:dupire_det_tiv}), respectively). Section \ref{sec:general_model}
further extends the model to allow drift and diffusion functions of general form
with arbitrary number of stochastic factors. The general setting is given by
(\ref{eqn:general_sde}) and the corresponding Dupire formula in
(\ref{eqn:generaldupire}). This general setup has greater coverage than the
examples found in literature as the interest rates are not assumed to follow
particular processes, such as short rate models; the assumption we have is that
the discount factor and the asset volatility are adapted functions of the It\^o
processes in the SDE system. An outline of the derivation for the vanilla call
option formulation with Hull-White short rates is presented in
\cite{Deelstra2012}. In contrast, we endeavor to present a complete and thorough
derivation for the general case. The rest of the section discusses specific
examples and implications for the leverage function in the stochastic local
volatility models as well as connections to Gy\"ongy's lemma. We chose to
present self-contained details of the derivations in the flow of the derivations
since that gives more insight into the actual foundation and possible
extensions. In Section 4, we demonstrate a calibration scheme with fixed-point
iteration for the local volatility model subject to stochastic rates, following
Linear Gaussian Model (LGM) processes; and study the convergence of the
iterations, and accuracy of the pricing with the calibrated local volatility
surface.

\section{Local Volatility Model with Two Stochastic Interest Rates}\label{sec:simple_model}

\subsection{Model Setup}\label{sec:initialsetup}

We assume the existence of a domestic risk neutral measure
$\mathbb{Q}^{\text{DRN}}$ that has the money market account $B^d_t$ as the
num\'eraire, which accrues at some computable short rate $r^d_t$ as $dB^d_t =
r^d_t B^d_t dt$. Dupire introduces a state dependent diffusion coefficient
$\sigma^{\text{LV}}(S_t, t)$ that uniquely describes the distribution of the
state variable $S_t$ for each time $t$, conditioned on the initial value $S_0$.
Accordingly, there is a risk-neutral spot process that is compatible with
observed market skew and allows a complete model. This is commonly
referred to as \emph{local volatility process} \cite{Dupire1994}, driven by the
Brownian motion $W^{S\text{(DRN)}}_t$ under the domestic risk neutral measure
$\mathbb{Q}^{\text{DRN}}$,
\begin{equation}
dS_t = \mu_t S_t dt
+ \sigma^{\text{LV}}(S_t, t) S_t dW^{S\text{(DRN)}}_t.\label{eqn:localvol}
\end{equation}
Dupire's formula gives the function
$\sigma^{\text{LV}}(\cdot, \cdot)$ in terms of call and/or put option
prices, or equivalently, implied volatility or total implied variance.

The original work of Dupire \cite{Dupire1994} assumes zero interest rates
($\mu_t = 0$), while the independent study of Derman and Kani
\cite{DermanKani1994} introduces deterministic interest rates ($\mu_t =
\mu(t)$).
In the latter setup, the drift term $\mu_t$ is assumed to be the instantaneous
forward rate of maturity $t$ implied from the yield curve, which is a
deterministic function of time. In this paper we relax this constraint and let
this term be stochastic. In particular, we are interested in a model with two
stochastic terms that comprise a drift of the form $\mu_t = \mu_t^1 - \mu_t^2$.
One can consider the pair $\mu_t^1$, $\mu_t^2$ as interest rate/dividend rate in
equities setup, or domestic rate/foreign rate in foreign exchange setup. In this
section, without loss of generality, we will use the conventions of the latter.
In particular, $\mu_t^1 = r^d_t$ and $\mu_t^2 = r^f_t$ denote the domestic and
foreign short rates, respectively. Under the domestic risk neutral measure
$\mathbb{Q}^{\text{DRN}}$, these rates follow single factor processes of the
generic form
\begin{equation}
\begin{split}
dr^d_t &= \alpha^d(\omega, t) dt + \sigma^d(\omega, t)
dW^{d\text{(DRN)}}_t,
\\
dr^f_t &= \alpha^f(\omega, t) dt + \sigma^f(\omega, t)
dW^{f\text{(DRN)}}_t,
\end{split}
\end{equation}
where $\alpha^d, \sigma^d, \alpha^f,\text{ and } \sigma^f$ are bounded
functions of a general set of stochastic factors $\omega$.
Our model admits three Brownian motions and we set the three pairs of
correlations\footnote{In general, the correlations can be time-dependent; or
they can even be generalized to stochastic processes as we shall see in
Section \ref{sec:general_model}. The nature of the correlations does not have
any impact on our result, thus we keep their notation simple.} as
$d\left<W^S, W^d\right>_t = \rho^{Sd} dt$, $\ \ d\left<W^S, W^f\right>_t =
\rho^{Sf} dt$, and $\ \ d\left<W^d, W^f\right>_t = \rho^{df} dt$. Note that the
short rate stochastic differential equations (SDEs) are typically written in the risk neutral measure
of their own currency. Here, drift adjustment in the foreign short rate process
due to the change from foreign risk neutral measure to domestic risk neutral
measure is absorbed into the term $\alpha^f$. When required
by our computations below, we assume that the functions $\sigma^{\text{LV}}$,
$\alpha^d$, $\sigma^d$, $\alpha^f$, and $\sigma^f$ are twice differentiable with
respect to the arguments over their entire ranges.

\subsection{Fokker-Planck equation}\label{sec:fokkerplanck}

Let $V_T \equiv V(S_T, r_T^d, r_T^f)$ be a twice differentiable $\mathbb{R}^3
\rightarrow \mathbb{R}$ test function.
For each given $T$, the discounted value of a European option maturing at $T$
with the payoff equal to that of the test function $V_T$ is a martingale under
the domestic risk neutral measure $\mathbb{Q}^{\text{DRN}}$,
\begin{equation}
\frac{V_0}{B_0^d} = V_0
= \mathbf{E}^{\mathbb{Q}^{\text{DRN}}}\left[\frac{V_T}{B_T^d}\right],
\end{equation}
where $B_T^d = \exp\left[\int_0^T r_u^d du\right]$ is the domestic money market
account, by means of which we can define $D_T \equiv \frac{1}{B_T^d}$ as the
corresponding discount factor.
In the (domestic) $T$-forward measure $\mathbb{Q}^{\text{T}}$, the zero coupon
bond $P^d(0, T)$ maturing at time $T$ is taken as the num\'{e}raire
\begin{equation}
P^d(t, T) = \mathbf{E}^{\mathbb{Q}^{\text{DRN}}} \left[\frac{D_T}{D_t}
\middle\vert \mathcal{F}_t \right], \text{and}\ 
\frac{d\mathbb{Q}^{\text{T}}}{d\mathbb{Q}^{\text{DRN}}} = 
\frac{D_T}{\mathbf{E}^{\mathbb{Q}^{\text{DRN}}}\left[D_T\right]} =
\frac{D_T}{P^d(0, T)}\label{eqn:dTdDRNRadonNikodym}
\end{equation}
where the filtration $\{\mathcal{F}_t, t \geq 0\}$ governs information arrival.
Thus
\begin{equation}
\begin{split}
\mathbf{E}^{\mathbb{Q}^{\text{DRN}}}\left[D_T V_T\right] &=
P^d(0, T)\mathbf{E}^{\mathbb{Q}^{\text{T}}}\left[V_T\right] \\
&= P^d(0, T) \int \int \int V(S_T, r^d_T, r^f_T)
\Phi^T(S_T, r^d_T, r^f_T, T) dS_T dr^d_T dr^f_T,\label{eqn:V0integral}
\end{split}
\end{equation}
where $\Phi^T(S_T, r^d_T, r^f_T, T)$ denotes the $T$-forward measure probability
density.
We assume the probability density function to be sufficiently tractable; in
particular, it is bounded; and it is differentiable with respect to time and
twice differentiable with respect to its spatial arguments. Notationwise, here
and in what is below, the integrals written without explicit limits are meant to
be taken over the entire domain, which is $(-\infty, \infty)$ in most cases.

One can integrate the full $T$-forward probability density $\Phi^T$
over the entire ranges of $r^d_T$ and $r^f_T$
to get the marginal $T$-forward probability density $q^T$ of $S_T$ over time,
\begin{equation}
q^T(S_T, T) = \int \int \Phi^T(S_T, r^d_T, r^f_T, T) dr^d_T dr^f_T.
\end{equation}
The marginal $T$-forward distribution has the time derivative
\begin{equation}
\frac{\partial q^T(S_T, T)}{\partial T} =
\int \int \frac{\partial \Phi^T(S_T, r^d_T, r^f_T, T)}{\partial T} dr^d_T
dr^f_T.
\end{equation}

Next, we apply It\^{o}'s lemma to the discounted test function,
\begin{equation*}
\begin{split}
\frac{d(D_T V_T)}{D_T} =&
\left[- r_T^d V_T
+ \frac{1}{2} (\sigma_T^{\text{LV}})^2 S_T^2 \frac{\partial^2 V_T}{\partial S_T^2}
+(r_T^d - r_T^f) S_T \frac{\partial V_T}{\partial S_T} \right.\\
&\left. + \frac{1}{2} (\sigma_T^d)^2 \frac{\partial^2 V_T}{\partial (r^d_u)^2}
+ \alpha^d_T \frac{\partial V_T}{\partial r^d_u}
+ \frac{1}{2} (\sigma_T^f)^2 \frac{\partial^2 V_T}{\partial (r^f_T)^2}
+ \alpha^f_T \frac{\partial V_T}{\partial r^f_T} \right.\\
&\left.
+\rho^{Sd} S_T \sigma_T^{\text{LV}} \sigma_T^d \frac{\partial^2 V_T}{\partial S_T \partial r^d_T}
+\rho^{Sf} S_T \sigma_T^{\text{LV}} \sigma_T^f \frac{\partial^2 V_T}{\partial S_T \partial r^f_T}
+\rho^{df} \sigma_T^d \sigma_T^f \frac{\partial^2 V_T}{\partial r^d_T \partial r^f_T} \right] dT\\
& + \sigma_T^{\text{LV}} S_T \frac{\partial V_T}{\partial S_T} dW_T^{S\text{(DRN)}}
+ \sigma_T^d \frac{\partial V_T}{\partial r^d_T} dW_T^{d\text{(DRN)}}
+ \sigma_T^f \frac{\partial V_T}{\partial r^f_T} dW_T^{f\text{(DRN)}}.
\end{split}
\end{equation*}
Here and below we use the convention $\sigma_T^{\text{LV}} \equiv
\sigma^{\text{LV}}(S_T, T)$, $\alpha_T^d \equiv \alpha^d(\omega, T)$,
$\sigma_T^d \equiv \sigma^d(\omega, T)$, $\alpha_T^f \equiv
\alpha^f(\omega, T)$, and $\sigma_T^f \equiv
\sigma^f(\omega, T)$ for notational brevity.
By taking the expectation in $\mathbb{Q}^{\text{DRN}}$, we find
\begin{equation}
\begin{split}
\frac{\partial\mathbf{E}^{\mathbb{Q}^{\text{DRN}}}\left[ D_T
V_T\right]}{\partial T} = \mathbf{E}^{\mathbb{Q}^{\text{DRN}}}\Big[ D_T \Big(& -
r_T^d V_T + \frac{1}{2} (\sigma_T^{\text{LV}})^2 S_T^2 \frac{\partial^2 V_T}{\partial S_T^2}
+(r_T^d - r_T^f) S_T \frac{\partial V_T}{\partial S_T} \\
&+ \frac{1}{2} (\sigma_T^d)^2 \frac{\partial^2 V_T}{\partial (r^d_T)^2}
+ \alpha^d_T \frac{\partial V_T}{\partial r^d_T}
+ \frac{1}{2} (\sigma_T^f)^2 \frac{\partial^2 V_T}{\partial (r^f_T)^2}
+ \alpha^f_T \frac{\partial V_T}{\partial r^f_T} \\
&+\rho^{Sd} S_T \sigma_T^{\text{LV}} \sigma_T^d \frac{\partial^2 V_T}{\partial
S_T \partial r^d_T} +\rho^{Sf} S_T \sigma_T^{\text{LV}} \sigma_T^f
\frac{\partial^2 V_T}{\partial S_T \partial r^f_T}\\
& +\rho^{df} \sigma_T^d
\sigma_T^f \frac{\partial^2 V_T}{\partial r^d_T \partial r^f_T} \Big)\Big]
\label{eqn:discountedEtimeder}
\end{split}
\end{equation}

On the other hand, we differentiate (\ref{eqn:V0integral}) with respect to $T$
to get
\begin{equation}
\begin{split}
\frac{\partial \left( P^d(0, T)\mathbf{E}^{\mathbb{Q}^{\text{T}}}\left[
V_T\right]\right)}{\partial T} =& \frac{\partial P^d(0, T)}{\partial T} \int
\int \int V_T \Phi^T dS_T dr^d_T dr^f_T\\
&+ P^d(0, T) \int \int \int 
V_T \frac{\partial \Phi^T}{\partial T} dS_T dr^d_T
dr^f_T,\label{eqn:dV0dT}
\end{split}
\end{equation}
By combining (\ref{eqn:discountedEtimeder}) and (\ref{eqn:dV0dT}) we
arrive at
\begin{align*}r0 =& \frac{\partial P^d(0, T)}{\partial T} \int \int \int V_T \Phi^T dS_T dr^d_T dr^f_T
+ P^d(0, T) \int \int \int V_T \frac{\partial \Phi^T}{\partial T} dS_T dr^d_T dr^f_T\\
& \begin{aligned}[t] + P^d(0, T) \int \int \int  \Phi^T
&\left[ r_T^d V_T - \frac{1}{2} (\sigma_T^{\text{LV}})^2 S_T^2 \frac{\partial^2 V_T}{\partial S_T^2}
-(r_T^d - r_T^f) S_T \frac{\partial V_T}{\partial S_T} \right.\\
&\left. - \frac{1}{2} (\sigma_T^d)^2 \frac{\partial^2 V_T}{\partial (r^d_T)^2}
- \alpha^d_T \frac{\partial V_T}{\partial r^d_T}
- \frac{1}{2} (\sigma_T^f)^2 \frac{\partial^2 V_T}{\partial (r^f_T)^2}
- \alpha^f_T \frac{\partial V_T}{\partial r^f_T} \right.\\
&\left.
-\rho^{Sd} S_T \sigma_T^{\text{LV}} \sigma_T^d \frac{\partial^2 V_T}{\partial S_T \partial r^d_T}
-\rho^{Sf} S_T \sigma_T^{\text{LV}} \sigma_T^f \frac{\partial^2 V_T}{\partial S_T
\partial r^f_T} \right.\\
&\left. 
-\rho^{df} \sigma_T^d \sigma_T^f \frac{\partial^2 V_T}{\partial r^d_T \partial r^f_T} \right] dS_T dr^d_T dr^f_T.
\end{aligned}
\end{align*}
Using the definition of the instantaneous forward rate
\begin{equation}
f^i(0, T) \equiv - \frac{\partial \log P^i(0, T) }{\partial T}
= - \frac{1}{P^i(0, T) }\frac{\partial P^i(0, T) }{\partial T},
\label{eqn:instfwdrate}
\end{equation}
with $i=d,f$, we can reformulate this as
\begin{equation}
\begin{split}
0 = \int \int \int \Bigg[
V_T \frac{\partial \Phi^T}{\partial T} + \Phi^T &\Bigg\{ \left(r_T^d -
f^d(0, T)\right) V_T \\
& - \frac{1}{2} (\sigma_T^{\text{LV}})^2 S_T^2
\frac{\partial^2 V_T}{\partial S_T^2} -(r_T^d - r_T^f) S_T \frac{\partial
V_T}{\partial S_T}\\
& - \frac{1}{2} (\sigma_T^d)^2 \frac{\partial^2 V_T}{\partial
(r^d_T)^2} - \alpha^d_T \frac{\partial V_T}{\partial r^d_T}
- \frac{1}{2} (\sigma_T^f)^2 \frac{\partial^2 V_T}{\partial (r^f_T)^2}
- \alpha^f_T \frac{\partial V_T}{\partial r^f_T} \\
&-\rho^{Sd} S_T \sigma_T^{\text{LV}} \sigma_T^d \frac{\partial^2 V_T}{\partial S_T \partial r^d_T}
-\rho^{Sf} S_T \sigma_T^{\text{LV}} \sigma_T^f \frac{\partial^2 V_T}{\partial S_T \partial r^f_T} \\
&-\rho^{df} \sigma_T^d \sigma_T^f \frac{\partial^2 V_T}{\partial r^d_T \partial r^f_T}
\Bigg\}
\Bigg] dS_T dr^d_T dr^f_T.\label{eqn:prefokkerplanck}
\end{split}
\end{equation}

The collection of zero coupon bond prices $P^i$ with maturities
sequenced over a time grid is called a \emph{discount curve}. The instantaneous
forward rates $f^i$ can be evaluated along given discount curves which are
used as standard input data in various pricing and other financial models.

We integrate by parts the terms that have the partial derivatives of $V_T$
appearing in (\ref{eqn:prefokkerplanck}). Noting that the boundary terms vanish
as we assume $\Phi^T$ and its derivatives tend to zero fast enough as its
arguments approach the integration limits, we can derive the following
identities by integrating by parts
\begin{equation*}
\begin{split}
\int \Phi^T f(\cdot) \frac{\partial^2 V_T}{\partial u^2} du =&
\int \frac{\partial^2 (\Phi^T f(\cdot))}{\partial u^2} V_T du,\\
\int \Phi^T f(\cdot) \frac{\partial V_T}{\partial u} du =&
-\int \frac{\partial (\Phi^T f(\cdot))}{\partial u} V_T du,\\
\int \Phi^T f(\cdot) \frac{\partial^2 V_T}{\partial u \partial v} du dv=&
\int \frac{\partial^2 (\Phi^T f(\cdot))}{\partial u \partial v} V_T du dv,
\end{split}
\end{equation*}
for a sufficiently well behaved (bounded, continuous, differentiable) function
$f$ of spatial coordinates $u$ and $v$, e.g. representing $S_T$, $r^d_T$,
$r^f_T$ in our setup.
Thus (\ref{eqn:prefokkerplanck}) can be written as
\begin{equation*}
\begin{split}
0 = \int \int \int V_T& \bigg\{
\frac{\partial \Phi^T}{\partial T} + \Phi^T \left(r_T^d - f^d(0, T)\right)
- \frac{1}{2} \frac{\partial^2 (\Phi^T(\sigma_T^{\text{LV}})^2 S_T^2)}{\partial S_T^2}
+(r_T^d - r_T^f) \frac{\partial (\Phi^T S_T)}{\partial S_T}\\
& - \frac{1}{2} \frac{\partial^2 (\Phi (\sigma_T^d)^2)}{\partial (r^d_T)^2}
+ \frac{\partial (\Phi^T \alpha^d_T)}{\partial r^d_T}
- \frac{1}{2} \frac{\partial^2 (\Phi^T (\sigma_T^f)^2)}{\partial (r^f_T)^2}
+ \frac{\partial (\Phi^T \alpha^f_T)}{\partial r^f_T} \\
& -\frac{\partial^2 (\Phi^T \rho^{Sd} S_T \sigma_T^{\text{LV}}
\sigma_T^d)}{\partial S_T \partial r^d_T} -\frac{\partial^2
(\Phi^T \rho^{Sf} S_T \sigma_T^{\text{LV}} \sigma_T^f)}{\partial S_T \partial
r^f_T}\\
& -\frac{\partial^2 (\Phi^T \rho^{df} \sigma_T^d \sigma_T^f)}{\partial
r^d_T \partial r^f_T} \bigg\}dS_T dr^d_T dr^f_T.
\end{split}
\end{equation*}
Since the above equation holds for any $V_T$, the term inside the braces
must vanish. This leads us to the \emph{Fokker-Planck (forward Kolmogorov)}
equation \cite{Feller1949}, which describes the evolution of the probability
density function $\Phi^T(S_T, r^d_T, r^f_T, T)$ of the underlying factors over
time,
\begin{equation}
\begin{split}
0 = &
\frac{\partial \Phi^T}{\partial T} + \Phi^T \left(r_T^d - f^d(0, T)\right)
- \frac{1}{2} \frac{\partial^2 (\Phi^T(\sigma_T^{\text{LV}})^2 S_T^2)}{\partial S_T^2}
+(r_T^d - r_T^f) \frac{\partial (\Phi^T S_T)}{\partial S_T}\\
& - \frac{1}{2} \frac{\partial^2 (\Phi (\sigma_T^d)^2)}{\partial (r^d_T)^2}
+ \frac{\partial (\Phi^T \alpha^d_T)}{\partial r^d_T}
- \frac{1}{2} \frac{\partial^2 (\Phi^T (\sigma_T^f)^2)}{\partial (r^f_T)^2}
+ \frac{\partial (\Phi^T \alpha^f_T)}{\partial r^f_T} \\
& -\frac{\partial^2 (\Phi^T \rho^{Sd} S_T \sigma_T^{\text{LV}}
\sigma_T^d)}{\partial S_T \partial r^d_T} -\frac{\partial^2 (\Phi^T \rho^{Sf} S_T
\sigma_T^{\text{LV}}\sigma_T^f )}{\partial S_T \partial r^f_T} -\frac{\partial^2
(\Phi^T \rho^{df} \sigma_T^d \sigma_T^f)}{\partial r^d_T \partial r^f_T}.
\label{eqn:fokkerplanck}
\end{split}
\end{equation}

\subsection{Extended Dupire formula}
\subsubsection{Call price surface formulation}\label{sec:DupireCall}

We integrate (\ref{eqn:fokkerplanck}) over the entire ranges of $r^d_T$ and
$r^f_T$. As before, as its arguments approach their limits the probability
distribution function $\Phi^T$ and its derivatives go to zero fast enough to
make the boundary terms vanish, and we obtain,
\begin{equation}
\begin{split}
0 =& \frac{\partial q^T}{\partial T} + \int \int \Phi^T \left(r_T^d - f^d(0, T)\right) dr^d_T dr^f_T
- \frac{1}{2} \frac{\partial^2 (q^T (\sigma_T^{\text{LV}})^2 S_T^2)}{\partial S_T^2}\\
&+ \frac{\partial}{\partial S_T} \left(\int \int (r_T^d - r_T^f) \Phi^T S_T dr^d_T dr^f_T\right).
\label{eqn:marginalfokkerplanck}
\end{split}
\end{equation}

The time zero value of a European vanilla call option $C$ with strike $K$,
which pays off $\max(S_T - K, 0)$ at time $T$ is given by
\begin{equation}
C = P^d(0, T) \mathbf{E}^{\mathbb{Q}^{\text{T}}}[(S_T-K) \mathds{1}_{S_T>K}]
= P^d(0, T) \int \int \int_K^{\infty} (S_T - K) \Phi^T dS_T dr^d_T dr^f_T
.\label{eqn:callexpectation}
\end{equation}
We compute the first two derivatives of the call price with respect to
strike $K$,
\begin{align}
\begin{split}
\frac{\partial C}{\partial K}
=& P^d(0, T) \int \int \left[-(S_T - K) \Phi^T \bigg\vert_{S_T=K}^{\infty}
-\int_K^{\infty} \Phi^T dS_T \right] dr^d_T dr^f_T\\
=& -P^d(0, T) \int \int \int_K^{\infty} \Phi^T dS_T dr^d_T dr^f_T
= -P^d(0, T) \mathbf{E}^{\mathbb{Q}^{\text{T}}}[\mathds{1}_{S_T > K}], \label{eqn:dCdK}
\end{split}\\
\begin{split}
\frac{\partial^2 C}{\partial K^2}
=& P^d(0, T) \int \int \Phi^T(K, r^d_T, r^f_T, T) dr^d_T dr^f_T
= P^d(0, T) q^T(K, T). \label{eqn:d2CdK2}
\end{split}
\end{align}
Next, we differentiate the call price with respect to time. Here we make use
of (\ref{eqn:marginalfokkerplanck}) and integration by parts,
\begin{align*}
\frac{\partial C}{\partial T}
=& \frac{\partial P^d(0, T)}{\partial T}
\int \int \int_K^{\infty} (S_T - K) \Phi^T dS_T dr^d_T dr^f_T\\
&+ P^d(0, T) \int \int \int_K^{\infty} (S_T - K) \frac{\partial \Phi^T}{\partial
T} dS_T dr^d_T dr^f_T\\
=& -f^d(0, T) C + P^d(0, T) \int_K^{\infty} (S_T - K) \frac{\partial q^T}{\partial T} dS_T\\
=& \begin{aligned}[t] -f^d(0, T) C + P^d(0, T) \int_K^{\infty} (S_T - K)
\bigg\{&
-\int \int \Phi^T \left(r_T^d - f^d(0, T)\right) dr^d_T dr^f_T\\
& + \frac{1}{2} \frac{\partial^2 (q^T (\sigma_T^{\text{LV}})^2 S_T^2)}{\partial S_T^2}\\
& - \frac{\partial}{\partial S_T} \left(\int \int (r_T^d - r_T^f) \Phi^T S_T dr^d_T dr^f_T\right)
\bigg\} dS_T
\end{aligned}\\
=& P^d(0, T) \int \int \int_K^{\infty} \Phi^T (K r_T^d - S_T r_T^f) dS_T dr^d_T dr^f_T
- \frac{1}{2} P^d(0, T) q^T S_T^2 (\sigma_T^{\text{LV}})^2 \bigg\vert_{S_T=K}^{\infty}.
\end{align*}
Plugging (\ref{eqn:d2CdK2}) into this expression yields
\begin{equation}
\frac{\partial C}{\partial T}
= P^d(0, T) \mathbf{E}^{\mathbb{Q}^{\text{T}}}\left[(K r_T^d - S_T r_T^f) \mathds{1}_{S_T > K}\right]
+ \frac{1}{2} K^2 \frac{\partial^2 C}{\partial K^2} (\sigma_T^{\text{LV}})^2 .
\end{equation}
Thus we arrive at the extended Dupire formula under stochastic rates
\begin{equation}
\sigma^{\text{LV}}(K, T)^2 = \frac{\frac{\partial C}{\partial T}
- P^d(0, T) \mathbf{E}^{\mathbb{Q}^{\text{T}}}\left[(K r_T^d - S_T r_T^f) \mathds{1}_{S_T > K}\right]}
{\frac{1}{2} K^2 \frac{\partial^2 C}{\partial K^2}}.\label{eqn:extendeddupire}
\end{equation}
This is an implicit formula, where the expectation on the right hand side
depends on the local volatility $\sigma^{\text{LV}}(K, T)$, and it can be
evaluated through a fixed-point iteration scheme. There is no known method to
compute the expectation above analytically, yet it can be evaluated by numerical
methods such as Monte Carlo or finite differences. Note also that the term with
the expectation corresponds to the price of an option with maturity $T$ and
payoff $(K r_T^d - S_T r_T^f) \mathds{1}_{S_T > K}$.

\paragraph{Single stochastic rate limit}

In the limit where the foreign rates $r_T^f$ are deterministic, this equation
becomes (see \cite{Hu2015} for an alternative derivation)
\begin{equation}
\sigma^{\text{LV}}(K, T)^2 = \frac{\frac{\partial C}{\partial T}
- P^d(0, T) K \mathbf{E}^{\mathbb{Q}^{\text{T}}}\left[r_T^d \mathds{1}_{S_T > K}\right]
+ P^d(0, T) r_T^f \mathbf{E}^{\mathbb{Q}^{\text{T}}}\left[S_T \mathds{1}_{S_T > K}\right]}
{\frac{1}{2} K^2 \frac{\partial^2 C}{\partial K^2}}.\label{eqn:extendeddupirereduced1}
\end{equation}
The second expectation in the numerator can be evaluated using
(\ref{eqn:callexpectation}) and (\ref{eqn:dCdK})
\begin{equation*}
P^d(0, T) r_T^f \mathbf{E}^{\mathbb{Q}^{\text{T}}}\left[S_T \mathds{1}_{S_T > K}\right]
= r_T^f \left[C + P^d(0, T) K \mathbf{E}^{\mathbb{Q}^{\text{T}}}[\mathds{1}_{S_T > K}] \right]
= r_T^f \left[C - K \frac{\partial C}{\partial K}\right],
\end{equation*}
which reduces (\ref{eqn:extendeddupirereduced1}) to
\begin{equation}
\sigma^{\text{LV}}(K, T)^2 = \frac{\frac{\partial C}{\partial T}
- P^d(0, T) K \mathbf{E}^{\mathbb{Q}^{\text{T}}}\left[r_T^d \mathds{1}_{S_T > K}\right]
+ r_T^f \left[C - K \frac{\partial C}{\partial K}\right]}
{\frac{1}{2} K^2 \frac{\partial^2 C}{\partial
K^2}}.\label{eqn:dupire_singlecall}
\end{equation}

\paragraph{Deterministic rates limit}

In the limit where both the domestic rates $r_T^d$ and the foreign rates $r_T^f$
are deterministic, one can evaluate the expectation in the above numerator using
(\ref{eqn:dCdK})
\begin{equation*}
- P^d(0, T) \mathbf{E}^{\mathbb{Q}^{\text{T}}}\left[\mathds{1}_{S_T > K}\right]
= \frac{\partial C}{\partial K}.
\end{equation*}
This allows us to reproduce the standard \emph{Dupire formula},
\begin{equation}
\sigma^{\text{LV}}(K, T)^2 = \frac{\frac{\partial C}{\partial T}
+ (r_T^d - r_T^f) K \frac{\partial C}{\partial K} + r_T^f C}
{\frac{1}{2} K^2 \frac{\partial^2 C}{\partial K^2}}.\label{eqn:dupire}
\end{equation}

\section{Generalized Local Volatility Model}\label{sec:general_model}

\subsection{Model Setup}\label{sec:generalizedsetup}

In (\ref{eqn:localvol}) we considered a standard local volatility process
of a particular form. Namely it is geometric and the drift term is a linear
combination of two stochastic rates, each modeled by a single factor process.
Here we relax these constraints and study the following general model with
drift and diffusion functions that allow arbitrary number of stochastic factors.

It is constructive to write down this SDE system in terms of $N$ independent
Brownian motions under the domestic risk neutral measure
$\mathbf{E}^{\mathbb{Q}^{\text{DRN}}}$,
\begin{equation*}
\begin{split}
dS_t =& \mu(\omega, t) dt +
   L(S_t, t) \sum_{k=1}^{N} \hat{\sigma}^S_k(\omega, t) d\hat{W}_t^{k},\\
dy^j_t =& \mu^j(\omega, t) dt +
   \sum_{k=1}^{N} \hat{\sigma}^j_k(\omega, t) d\hat{W}_t^{k},\\
d\left<\hat{W}^{k}, \hat{W}^{l}\right>_t =& \delta^{kl} dt,
\end{split}
\end{equation*}
where $\mu, \hat{\sigma}^S_k, \mu^j, \text{ and } \hat{\sigma}^j_k$ are
bounded functions of a general set of stochastic factors $\omega$.
$Y_t \equiv (y^1_t, \ldots, y^M_t)$ is the set of additional It\^{o} processes
in the SDE system for which we do not assume any special form other than the
above. $L(S_t, t)$ is the local volatility or \emph{leverage function} we want
to compute.
In this setup, we observe that the correlation structure of the underlying
assets is absorbed into the functions $\hat{\sigma}^S_k$ and
$\hat{\sigma}^j_k$. The correlations themselves can be It\^{o}
processes, in which case they are assigned to particular $y^j_t$s.
The (domestic) discount factor $D_t$ is an adapted function of
$Y_t$; yet in general we do not assume a particular mapping \footnote{In the
foreign exchange setting of Section \ref{sec:simple_model}, $r_t^d$ and $r_t^f$
are each direct components of $Y_t$. We will return to this particular case in
Section \ref{sec:examples}. As another example, in case $r_t^d$ follows a
multi-factor short rate model, it can be written as a function of the factors
that are a subset of $Y_t$.}.

The SDE system can also be written in terms of correlated Brownian motions
split into those driving the processes of $S_t$ and $Y_t$ separately, with $N =
N_S + N_Y$, as
\begin{equation}
\begin{split}
dS_t =& \mu(\omega, t) dt +
   L(S_t, t) \sum_{k=1}^{N_S} \sigma^S_k(\omega, t) dW_t^{Sk},\\
dy^j_t =& \mu^j(\omega, t) dt +
   \sum_{k=1}^{N_Y} \sigma^j_k(\omega, t) dW_t^{Yk}.
\end{split} \label{eqn:general_sde}
\end{equation}

\subsection{Fokker-Planck Equation}

Following the same methodology from Section
\ref{sec:fokkerplanck}, omitting repetitive parts of the computation, we derive
the corresponding Fokker-Planck equation. We compute the time
derivative of the discounted value of a twice differentiable test function $V_T
= V(S_T, Y_T)$ as
\begin{equation}
\begin{split}
\frac{\partial\mathbf{E}^{\mathbb{Q}^{\text{DRN}}}\left[ D_T
V_T\right]}{\partial T} = \mathbf{E}^{\mathbb{Q}^{\text{DRN}}}\Big[ D_T \Big(&
- r_T^d V_T
+ \frac{1}{2} L_T^2 \bar{\sigma}_T^2 \frac{\partial^2 V_T}{\partial S_T^2}
+\mu_T \frac{\partial V_T}{\partial S_T}\\
&+ \text{terms involving $Y_T$ derivatives of $V_T$}\Big)\Big],
\end{split}
\end{equation}
where $r_T^d \equiv -\frac{1}{D_T}\frac{\partial D_T}{\partial T}$ is not
assumed to follow a particular stochastic process, we defined
\begin{equation*}
\bar{\sigma}_T^2 \equiv \sum_{l,m=1}^{N_S}
\left({\sigma}^S_l \rho^S_{lm} {\sigma}^S_m
\right)
= \sum_{k=1}^{N}
\left(\hat{\sigma}^S_k \right)^2,
\end{equation*} and to keep the notation compact we
denoted $\mu_T = \mu(\omega, T)$ and $L_T = L(S_T, T)$. Here $\rho^S_{lm}$
denotes the correlation function between the Brownian motions $W_t^{Sl}$, i.e.
$d\left<W^{Sl}, W^{Sm}\right>_t = \rho^S_{lm} dt$. Combining this with the
expression for the time derivative of the discounted value of the test function
in $T$-forward measure $\mathbb{Q}^{\text{T}}$ yields
\begin{equation}
\begin{split}
0 =& \frac{\partial P^d(0, T)}{\partial T} \int \int V_T \Phi^T dS_T dY_T
+ P^d(0, T) \int \int V_T \frac{\partial \Phi^T}{\partial T} dS_T dY_T\\
& \begin{aligned}[t] + P^d(0, T) \int \int  \Phi^T
&\Big{[} r_T^d V_T
- \frac{1}{2} L_T^2 \bar{\sigma}_T^2 \frac{\partial^2 V_T}{\partial S_T^2}
-\mu_T \frac{\partial V_T}{\partial S_T} \\
& + \text{terms involving $Y_T$ derivatives of $V_T$} \Big{]} dS_T
dY_T,\label{eqn:dPVdT}
\end{aligned}
\end{split}
\end{equation}
where $\Phi^T(S_T, Y_T, T)$ is the $T$-forward measure
probability density, with the corresponding marginal density
$q^T(S_T, T) = \int \Phi^T(S_T, Y_T, T) dY_T$, which can be used to
formulate the derivatives of the call option price with respect to strike,
analogous to (\ref{eqn:dCdK}) and (\ref{eqn:d2CdK2}) as
\begin{equation*}
\begin{split}
\frac{\partial C}{\partial K}
=& -P^d(0, T) \mathbf{E}^{\mathbb{Q}^{\text{T}}}[\mathds{1}_{S_T > K}],\\
\frac{\partial^2 C}{\partial K^2}
=& P^d(0, T) q^T(K, T).
\end{split}
\end{equation*}
Moreover, to elucidate the notation, we make a note that the integrals along
the stochastic factors $Y_T$
\begin{equation*}
\int f(\cdot) dY_T \equiv \int \ldots \int f(\cdot) dy_T^1 dy_T^2 \ldots
\end{equation*}
are taken over their entire domains.

Since $\frac{\partial P^d(0, T)}{\partial T} = -f^d(0, T)P^d(0, T)$,
(\ref{eqn:dPVdT}) becomes
\begin{equation*}
\begin{split}
0 = \int \int \Big{[}
V_T \frac{\partial \Phi^T}{\partial T} + \Phi^T &\Big{\{} \left(r_T^d - f^d(0, T)\right) V_T
- \frac{1}{2} L_T^2 \bar{\sigma}_T^2 \frac{\partial^2 V_T}{\partial S_T^2}
-\mu_T \frac{\partial V_T}{\partial S_T} \\
& + \text{terms involving $Y_T$ derivatives of $V_T$}
\Big{\}}
\Big{]} dS_T dY_T.
\end{split}
\end{equation*}
As before, we integrate by parts the above integrals to factor out $V_T$. This
leads to the following Fokker-Planck equation,
\begin{equation}
\begin{split}
0 = &
\frac{\partial \Phi^T}{\partial T} + \Phi^T \left(r_T^d - f^d(0, T)\right)
- \frac{1}{2} \frac{\partial^2 (\Phi^T L_T^2 \bar{\sigma}_T^2)}{\partial S_T^2}
+ \frac{\partial (\Phi^T \mu_T)}{\partial S_T}\\
& + \text{terms involving $Y_T$ derivatives of $\Phi^T$}.
\label{eqn:fokkerplanck_general}
\end{split}
\end{equation}

\subsection{Generalized Dupire formula}
As in Section \ref{sec:DupireCall}, we integrate the Fokker-Planck equation
(\ref{eqn:fokkerplanck_general}) over the entire ranges of $Y_T$.
The probability distribution function $\Phi^T$ goes to zero fast enough as
its arguments approach their limits, making the boundary terms that involve
the $Y_T$ derivatives vanish,
\begin{equation}
\begin{split}
0 =& \frac{\partial q^T}{\partial T}
+ \int \Phi^T \left(r_T^d - f^d(0, T)\right) dY_T
- \frac{1}{2} \frac{\partial^2}{\partial S_T^2}
\left(L_T^2 \int \Phi^T \bar{\sigma}_T^2 dY_T \right)\\
&+ \frac{\partial}{\partial S_T} \left(\int \Phi^T \mu_T dY_T \right).
\label{eqn:marginalfokkerplanck_general}
\end{split}
\end{equation}
At this point we note that the terms involving the correlation coefficients are
all integrated out, therefore we conclude that the nature of the correlations
will not have any impact on our result.
Next we compute the time derivative of the price of a European vanilla call
option $C$ with strike $K$. Here we make use of the definition of conditional
expectation,
$
\Phi^T(Y, T |S_T=X) \equiv \frac{\Phi^T(X, Y, T)}{q^T(X, T)},
$
as well as
(\ref{eqn:marginalfokkerplanck_general}) and integration by parts,
\begin{align*}
\frac{\partial C}{\partial T}
=& \frac{\partial P^d(0, T)}{\partial T} \int \int_K^{\infty} (S_T - K) \Phi^T dS_T dY_T
+ P^d(0, T) \int \int_K^{\infty} (S_T - K) \frac{\partial \Phi^T}{\partial T} dS_T dY_T\\
=& -f^d(0, T) C + P^d(0, T) \int_K^{\infty} (S_T - K) \frac{\partial q^T}{\partial T} dS_T\\
=& \begin{aligned}[t] -f^d(0, T) C + P^d(0, T) \int_K^{\infty} (S_T - K)
\bigg\{&
-\int \Phi^T \left(r_T^d - f^d(0, T)\right) dY_T\\
& + \frac{1}{2} \frac{\partial^2}{\partial S_T^2}
\left(L_T^2 \int \Phi^T \bar{\sigma}_T^2 dY_T\right)\\
& - \frac{\partial}{\partial S_T} \left(\int \Phi^T \mu_T dY_T\right)
\bigg\} dS_T
\end{aligned}\\
=& P^d(0, T) \int \int_K^{\infty} \Phi^T \left[\mu_T - (S_T - K) r_T^d \right]
dS_T dY_T\\
& + \frac{1}{2} P^d(0, T) q^T(K, T) L(K, T)^2 \int \int \Phi^T(Y_T, T | S_T=K)
\bar{\sigma}_T^2 dY_T\\
=& P^d(0, T) \mathbf{E}^{\mathbb{Q}^{\text{T}}}
\left[\left\{\mu_T - (S_T - K) r_T^d\right\} \mathds{1}_{S_T > K} \right]
+ \frac{1}{2} L(K, T)^2 \frac{\partial^2 C}{\partial K^2}
\mathbf{E}^{\mathbb{Q}^{\text{T}}}\left[\bar{\sigma}_T^2 | S_T=K\right].
\end{align*}
This gives us the generalized form of the Dupire formula,
\begin{equation}
L(K, T)^2 = \frac{\frac{\partial C}{\partial T}
- P^d(0, T) \mathbf{E}^{\mathbb{Q}^{\text{T}}}
\left[\left\{\mu_T - (S_T - K) r_T^d\right\} \mathds{1}_{S_T > K} \right]}
{\frac{1}{2} \frac{\partial^2 C}{\partial K^2}
\mathbf{E}^{\mathbb{Q}^{\text{T}}}\left[\bar{\sigma}_T^2 | S_T=K\right]}.\label{eqn:generaldupire}
\end{equation}
As was the case with the extended Dupire formula (\ref{eqn:extendeddupire}),
this is an implicit equation, where the expectations on the right hand side
depend on the leverage function $L(K, T)$, and it can be evaluated through a
fixed-point iteration scheme.

\subsection{Examples}\label{sec:examples}

The generalized Dupire formula (\ref{eqn:generaldupire}) applies to a wide
range of models.

\subsubsection{Simple Models}

For simplicity, we consider the underlier $S_t$ to be driven by a single
Brownian motion ($N_S=1$) in this section,
\begin{equation*}
dS_t = \mu(\omega, t) dt +
   L_s(S_t, t) \bar{\sigma}(\omega, t) dW_t^{S},
\end{equation*}
where $L_s$ denotes the leverage function of the
simplified model, to distinguish it from the generalized model.

The special case of this model with $\bar{\sigma} = S_t$ is of
special interest where the SDE becomes a simple local volatility model. In this case
(\ref{eqn:generaldupire}) becomes
\begin{equation}
\sigma^{\text{LV}}(K, T)^2 = \frac{\frac{\partial C}{\partial T}
- P^d(0, T) \mathbf{E}^{\mathbb{Q}^{\text{T}}}
\left[\left\{\mu_T - (S_T - K) r_T^d\right\} \mathds{1}_{S_T > K} \right]}
{\frac{1}{2} K^2 \frac{\partial^2 C}{\partial K^2}}.
\label{eqn:generaldupire_simplified}
\end{equation}
Comparison of (\ref{eqn:generaldupire}) with
(\ref{eqn:generaldupire_simplified}) gives us the following relationship between
the generalized model and its corresponding simple local volatility
simplification,
\begin{equation}
\sigma^{\text{LV}}(K, T)^2 K^2
= L(K, T)^2 \mathbf{E}^{\mathbb{Q}^{\text{T}}}\left[\bar{\sigma}^2 |
S_T=K\right].
\end{equation}

To recover the simpler FX local volatility model with two stochastic rates
($ Y_t = r^d_t, r^f_t$) from Section \ref{sec:initialsetup}, one can set
$\mu_t = (r_t^d - r_t^f) S_t$ and $\bar{\sigma} = S_t$. In this case it is
straightforward to show that (\ref{eqn:generaldupire}) reduces to
(\ref{eqn:extendeddupire}).

\subsubsection{Stochastic Local Volatility}

An extension of the simple model is the \emph{stochastic local volatility (SLV)}
model with $\bar{\sigma} = S_t \sqrt{U_t}$ where $U_t$ is the variance process,
\begin{equation*}
dU_t = \mu^U(U_t, t) dt + \sigma^U(U_t, t) dW^U_t,
\end{equation*}
which is typically chosen to fit certain options or ranges or aspects of the
price, and the leverage function  $L(S_t, T)$ serves as a correction that
ensures that all vanilla options are repriced over the full calibration range. A
common choice is to use a Cox-Ingersoll-Ross (CIR) process \cite{CIR1985}
in which case in the context of FX derivatives the SDE system becomes
\cite{Tian2012,Tataru2012},
\begin{equation}
\begin{split}
dS_t =& (r_t^d - r_t^f) S_t dt + L_s(S_t, t) S_t \sqrt{U_t}
dW^{S\text{(DRN)}}_t\\
dU_t =& \kappa(\theta - U_t) dt + \xi \sqrt{U_t} dW^{U\text{(DRN)}}_t,
\end{split}\label{eqn:FXslvSDE}
\end{equation}
where mean reversion speed $\kappa$, long term mean $\theta$, and vol-of-vol
$\xi$ are possibly time-dependent CIR parameters.

For this model, the generalized Dupire formula (\ref{eqn:generaldupire})
simplifies to
\begin{equation}
L_s(K, T)^2 = \frac{\frac{\partial C}{\partial T}
- P^d(0, T) \mathbf{E}^{\mathbb{Q}^{\text{T}}}
\left[(K r_T^d - S_T r_T^f) \mathds{1}_{S_T > K}\right]}
{\frac{1}{2} K^2 \frac{\partial^2 C}{\partial K^2}
\mathbf{E}^{\mathbb{Q}^{\text{T}}}\left[U_T | S_T=K\right]}.\label{eqn:dupire_slv}
\end{equation}
Here we emphasize that the above equation assumes only the particular form of
the SDE for the underlier $S_t$, and is not restricted to the case where the SDE
for the variance $U_t$ is of type CIR.

Comparing (\ref{eqn:extendeddupire}) to (\ref{eqn:dupire_slv}) allows us to
write the relationship between the local volatility function of the simple local
volatility model (\ref{eqn:localvol}) and the leverage function of the
stochastic local volatility model (\ref{eqn:FXslvSDE}) as
\begin{equation}
\sigma^{\text{LV}}(K, T)^2
= L_s(K, T)^2 \mathbf{E}^{\mathbb{Q}^{\text{T}}}\left[U_T |
S_T=K\right].\label{eqn:linkLV2SLV}
\end{equation}

This relationship is reached independently from but is consistent with
Gy\"{o}ngy's finding \cite{Gyongy1986} that links the set of stochastic
processes $X_t \equiv \{X_t^i\}$ with It\^{o} differentials
\begin{equation*}
dX_t^m = \alpha^m(t, \omega) dt + \sum_{n=1}^{N} \beta^{mn}(t, \omega)
d\tilde{W}_t^n,
\end{equation*}
where $\alpha^m$ and $\beta^{mn}$ are bounded functions of a general set of
stochastic factors $\omega$, which may include factors not contained in or related to
$X_t$, and $\{\tilde{W}_t^i\}$ are Brownian motions under measure $\mathbb{P}$;
to another set of stochastic processes $Z_t \equiv \{Z_t^i\}$ with deterministic
coefficients $a^m$ and $b^{mn}$,
\begin{equation*}
dZ_t^m = a^m(t, Z_t) dt + \sum_{n=1}^{N} b^{mn}(t, Z_t)
d\hat{W}_t^n,
\end{equation*}
where $\{\hat{W}_t^i\}$ are Brownian motions under measure $\mathbb{Q}$,
in that the two sets of processes have the same marginal probability
distribution, $X_t$ under $\mathbb{P}$ and of $Z_t$ under
$\mathbb{Q}$, for every $t$ if
\begin{align*}
a^m(t, z) =& \mathbf{E}^\mathbb{P}\left[\alpha^m(t, \omega) \arrowvert X_t =
z\right], \\
\sum_{p=1}^N b^{mp}(t, z) b^{np}(t, z)  =&
\mathbf{E}^\mathbb{P}\left[\sum_{p=1}^N \beta^{mp}(t, \omega) \beta^{np}(t,
\omega) \arrowvert X_t = z\right],
\end{align*}
for all $m, n = 1, \ldots, N$.

Applying this to our example, the marginal distribution of the stochastic local
volatility model (\ref{eqn:FXslvSDE}) must be the same as the
distribution of the simple local volatility model (\ref{eqn:localvol}) if
(\ref{eqn:linkLV2SLV}) holds.
This implies that having computed the function $\sigma^{\text{LV}}(K, T)$
for the simple local volatility model using
(\ref{eqn:extendeddupire}), one can obtain the leverage function $L_s(K, T)$ of
the stochastic local volatility model by evaluating the conditional expectation
$\mathbf{E}^{\mathbb{Q}^{\text{T}}}\left[U_T | S_T=K\right]$. 
One utilizes a numerical method such as multi-dimensional finite difference or
Monte Carlo simulation to estimate this conditional expectation as there is no
straightforward way to evaluate it analytically.

\section{Case study}

In this section we study FX local volatility model where the domestic and
foreign rates are governed by stochastic processes. For the rates evolution we
consider the Linear Gaussian Model, which we first briefly introduce.

\subsection{Linear Gaussian Model}
In the Linear Gaussian Model (LGM), as proposed by \cite{HW1999}, the
rate is driven by a single Markovian factor
\begin{equation*}
dx_t = \sigma_t dW^N_t,\ x_0 = 0,
\end{equation*}
in the measure $\mathbb{Q}^{\text{N}}$ defined by the num\'eraire
\begin{equation*}
N(t, x_t) = \frac{1}{P(0, t)} \exp\left[H_t x_t + \frac{1}{2} H^2_t
\zeta_t\right],
\end{equation*}
where
\begin{eqnarray*}
\zeta_t &= \int_0^t \sigma^2_s ds,\\
H_t &= \int_0^t h_s ds,
\end{eqnarray*}
and the model parameters $\sigma_t$ and $h_t$ are calibrated to the market
quotes. The short rate is given by \cite{LSG2015}
\begin{equation*}
r(t, x_t) = f(0, t) + h_t x_t + h_t H_t \zeta_t,
\end{equation*}
where the instantaneous forward rate $f(0, t)$ is computed from the zero
coupon bond curve $P(0, t)$ as in (\ref{eqn:instfwdrate}).

The change to the risk neutral measure $\mathbb{Q}^{\text{RN}}$, where the money
market account is the num\'eraire, is derived as \cite{HW1999},
\begin{equation*}
dW^{\text{(RN)}}_t = dW^N_t + \sigma_t H_t dt,
\end{equation*}
so that the Markovian factor evolves in this measure as
\begin{equation*}
dx_t = - \sigma^2_t H_t dt + \sigma_t dW^{\text{(RN)}}_t.
\end{equation*}

\subsection{Local Volatility with LGM rates}
We consider a three-factor local volatility model for the FX rate $S$, where the
domestic rate $r^d$ and foreign rate $r^f$ are governed by LGM processes, with
time dependent parameters $\sigma^d, h^d$ and $\sigma^f, h^f$ respectively.
The joint evolution of the three factors is given by
\begin{equation}
\begin{split}
dS_t =& \left[r_t^d - r_t^f \right] S_t dt + \sigma^{\text{LV}}(S_t, t) S_t
dW^{S\text{(DRN)}}_t,\\
dx^d_t =& - (\sigma^d_t)^2 H^d_t dt + \sigma^d_t dW^{d\text{(DRN)}}_t,\\
dx^f_t =& - \left[(\sigma^f_t)^2 H^f_t + \rho^{Sf} \sigma^f_t
\sigma^{\text{LV}}(S_t, t) \right] dt +
\sigma^f_t dW^{f\text{(DRN)}}_t.\\
\end{split}\label{eqn:LV2SR_LGM}
\end{equation}
The coefficients of correlation between the Brownian motions
$W^{S\text{(DRN)}}_t$, $W^{d\text{(DRN)}}_t$, $W^{f\text{(DRN)}}_t$, evolving
under the domestic risk neutral measure
$\mathbf{E}^{\mathbb{Q}^{\text{DRN}}}$, are given by the quadratic covariances
\begin{equation*}
\begin{split}
d\left<W^{S\text{(DRN)}}, W^{d\text{(DRN)}} \right>_t = & \rho^{Sd} dt,\\
d\left<W^{S\text{(DRN)}}, W^{f\text{(DRN)}} \right>_t = & \rho^{Sf} dt,\\
d\left<W^{d\text{(DRN)}}, W^{f\text{(DRN)}} \right>_t = & \rho^{df} dt.
\end{split}
\end{equation*}

\subsection{Calibrating the local volatility surface}

The Radon-Nikodym derivative (\ref{eqn:dTdDRNRadonNikodym})
allows us to transform the extended Dupire formula (\ref{eqn:extendeddupire}) to
\begin{equation}
\sigma^{\text{LV}}(K, T)^2 = \frac{\frac{\partial C}{\partial T}
- \mathbf{E}^{\mathbb{Q}^{\text{DRN}}}\left[D_T (K r_T^d - S_T r_T^f)
\mathds{1}_{S_T > K}\right]} {\frac{1}{2} K^2 \frac{\partial^2 C}{\partial
K^2}}.\label{eqn:extendeddupire_RN}
\end{equation}
The expectation in the above expression can be estimated by Monte Carlo
simulation.

We propose the following algorithm to calibrate the local volatility surface
at time slices $t_j, j = 1, \ldots, M$, by solving (\ref{eqn:extendeddupire_RN})
in a fixed point iteration scheme:
\begin{enumerate}
  \item Using the market implied volatility $\Sigma(K, t)$, generate a vanilla
  call option price surface $C(K, t)$ interpolator (or a total implied variance
  surface $w(y, t)$ interpolator).
  \item For the first time slice $T=t_1 > 0$, \textbf{(a)} in the first
  iteration evaluate the deterministic equation (\ref{eqn:dupire}) to compute
  the FX local volatilities for a predetermined range of strikes. This step
  requires no Monte Carlo simulation.
   \textbf{(b)} in the subsequent iterations, simulate the SDE system
   (\ref{eqn:LV2SR_LGM}) up to time $t_1$ using the local vol values from
   previous iteration. Compute the Monte Carlo
  estimate for the expectation $\mathbf{E}^{\mathbb{Q}^{\text{DRN}}}\left[D_T (K r_T^d - S_T r_T^f)
\mathds{1}_{S_T > K}\right]$ appearing in (\ref{eqn:extendeddupire_RN})
  for the same set of strikes. Update local vol values with this equation.
  \item For each of the subsequent time slices $T=t_j, j > 1$, simulate the SDE
  system (\ref{eqn:LV2SR_LGM}) up to time $t_j$, where \textbf{(a)} in the first
  iteration use the local vol values from time slice $t_{j - 1}$ for time slice
  $t_j$, \textbf{(b)} in the subsequent iterations use local vol values from
  previous iteration for time slice $t_j$; and linearly interpolate the local
  volatility values between $t_{j - 1}$ and $t_j$ slices for simulation times
  between the slices. Compute the
  Monte Carlo estimate for the expectation $\mathbf{E}^{\mathbb{Q}^{\text{DRN}}}\left[D_T (K r_T^d - S_T r_T^f)
\mathds{1}_{S_T > K}\right]$ for a predetermined range of
  strikes.
  Update local vol values with the extended Dupire equation
  (\ref{eqn:extendeddupire_RN}).
\end{enumerate}

\subsection{Implementation and Results}

The EUR-USD market data as of 2021-09-30 we use in our analysis contains the
EUR-USD spot FX rate $S_0$, the EUR-USD implied volatility surface representing
market quotes of vanilla option instruments on FX rate $S_t$, the discount
curves $P^d(0, t)$, $P^f(0, t)$ for the domestic rate $r^d_t$ and the foreign
rate $r^f_t$, respectively. The coefficients of correlation are given by
$\rho^{Sd} = 0.059, \rho^{Sf} = 0.031, \rho^{df} = 0.255$. For LGM model
parameters, we use $\sigma^d_t = \sigma^f_t = 0.01$, and $h^d_t = h^f_t = 1$.

\begin{figure}[ht!]
    \centering
    \includegraphics[width=\textwidth]{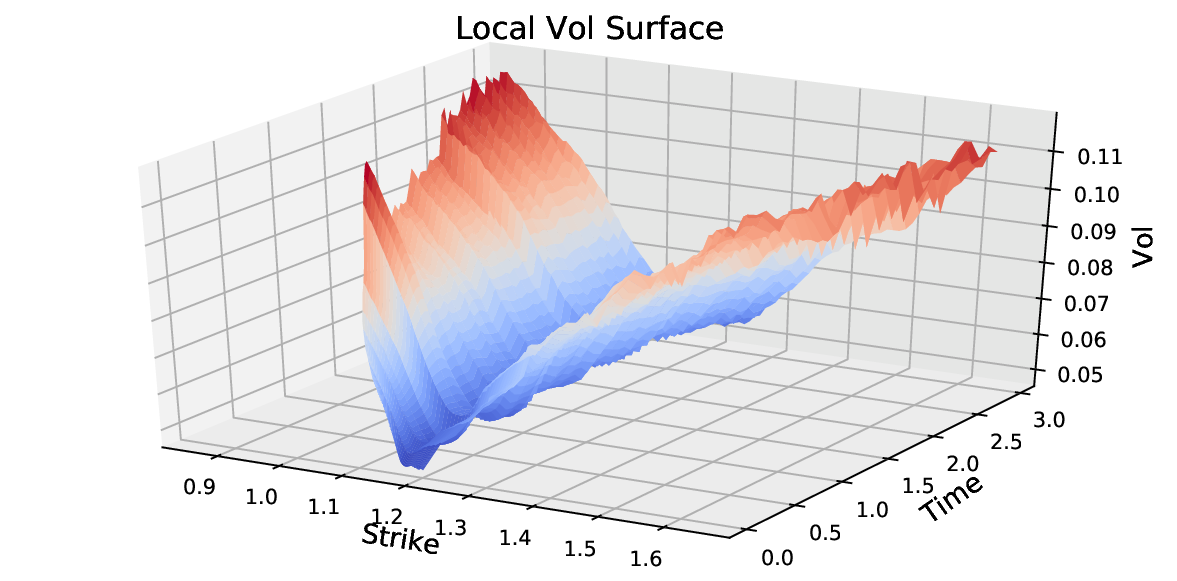}
    \caption{Calibrated EUR-USD local volatility}
    \label{fig:calibratedLocalVol}
\end{figure}

We calibrate the local volatility at time slices $t_j = 0.05, 0.1, \ldots, 3.0$.
At each time slice, we select the strike grid $K_j^l$ to span the strike range
$[F_{t_j} e^{-3 \Sigma(F_{t_j}, t_j) \sqrt{t_j}}, F_{t_j}
e^{3 \Sigma(F_{t_j}, t_j) \sqrt{t_j} }]$ uniformly spaced in log-moneyness,
where $F_{t_j} = S_0 \frac{P^f(0,t_j)}{P^d(0,t_j)}$ is the forward asset price,
and $\Sigma(F_{t_j}, t_j)$ is the at-the-money-forward implied volatility.

\begin{figure}[ht!]
    \centering
    \includegraphics[width=\textwidth]{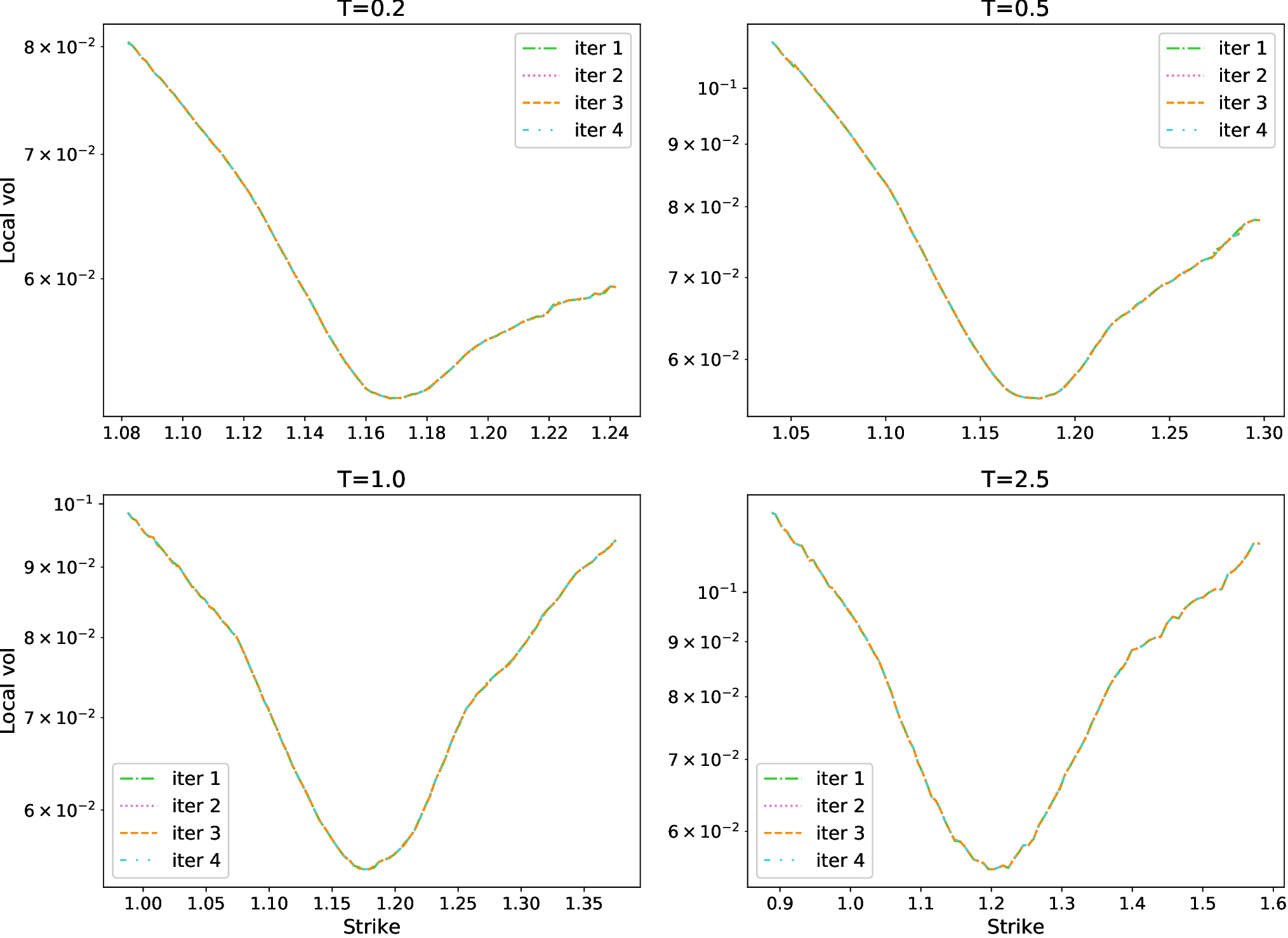}
    \caption{Calibrated local volatility values at several time slices after
    each iteration}
    \label{fig:localVolIterations}
\end{figure}

The setup for simulation is as follows. The simulation time step size is set to
0.004 years, so that the simulation times are $t_i = 0.004, 0.008, \ldots$. The
SDE system (\ref{eqn:LV2SR_LGM}) is simulated up to calibration time $t_j$ in
forward Euler scheme,
\begin{equation}
\begin{split}
\Delta S_{i + 1} =& \left[r_i^d - r_i^f \right] S_i \Delta t_i +
\sigma^{\text{LV}}(S_i, t_i) S_i \sqrt{\Delta t_i} Z^S_i,\\
\Delta x^d_{i + 1} =& - (\sigma^d_{t_i})^2 H^d_{t_i} \Delta t_i + \sigma^d_{t_i}
\sqrt{\Delta t_i} Z^d_i,\\
\Delta x^f_{i + 1} =& - \left[(\sigma^f_{t_i})^2 H^f_{t_i} + \rho^{Sf}
\sigma^f_{t_i} \sigma^{\text{LV}}(S_i, t_i) \right] \Delta t_i +
\sigma^f_{t_i} \sqrt{\Delta t_i} Z^f_i,\\
\end{split}\label{eqn:LV2SR_LGM_forwardEuler}
\end{equation}
where $\Delta t_i = t_{i + 1} - t_i$. The random numbers $(Z^S_i, Z^d_i, Z^f_i)$
are drawn from the normal distribution $N(0, \Sigma)$, where
\begin{equation*}
\Sigma = 
\begin{pmatrix}
1 & \rho^{Sd} & \rho^{Sf}\\
\rho^{Sd} & 1 & \rho^{df}\\
\rho^{Sf} & \rho^{df} & 1
\end{pmatrix},
\end{equation*}
using Cholesky decomposition.

\begin{figure}[ht!]
    \centering
    \includegraphics[width=\textwidth]{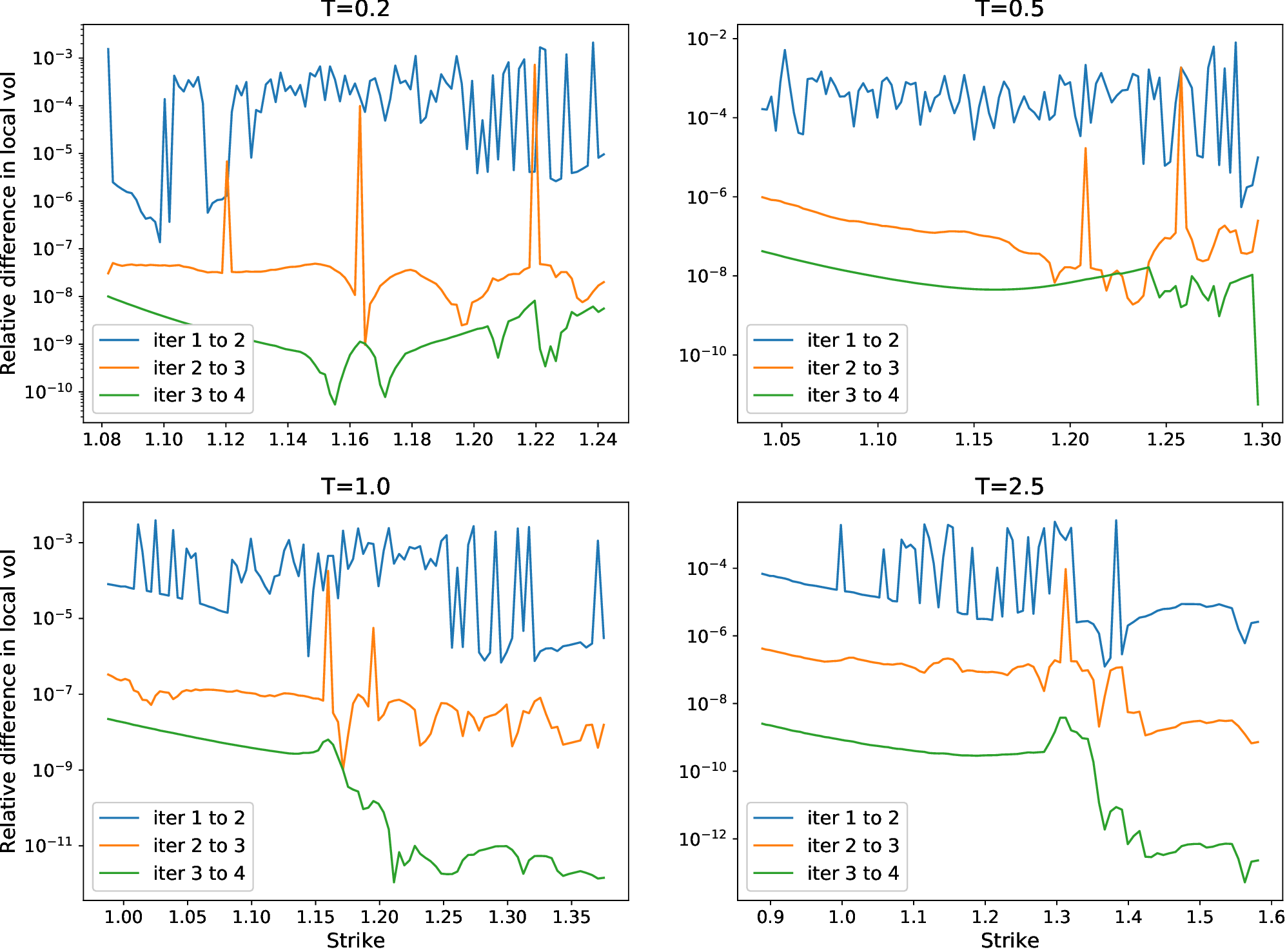}
    \caption{The local volatility update
    $\frac{|\sigma^{\text{LV}}_{I+1} -
    \sigma^{\text{LV}}_I|}{\sigma^{\text{LV}}_I}$ after iterations $I =
    1, 2, 3$, at several time slices}
    \label{fig:relativeDiffLocalVol}
\end{figure}

We simulate 1000 paths and their
antithetic conjugate to compute the Monte Carlo average of
$\left[D_{t_j} (K^l_{t_j} r_{t_j}^d - S_{t_j} r_{t_j}^f)
\mathds{1}_{S_{t_j} > K^l_{t_j}}\right]$ for all $K^l_{t_j}$ in the strike grid
at time slice $t_j$, which we plug in to the expectation in
(\ref{eqn:extendeddupire_RN}) to compute $\sigma^{\text{LV}}(K^l_{t_j}, t_j)$.
The number of iterations at each time slice is set to 4.

In Figure \ref{fig:calibratedLocalVol} the surface plot can be seen as a
visual evidence that the local volatility calibration completed smoothly. To see
the impact of successive iterations, we plot the local volatility curve after
each iteration at time slices $t_j = 0.2, 0.5, 1.0, 2.5$ in Figure
\ref{fig:localVolIterations}. As can be seen in the plots, the successive
iterations do not result in a visual change in the calibrated local volatility
values. In Figure \ref{fig:relativeDiffLocalVol} we see that the relative
differences in local volatility between successive iterations shrinks fast,
which shows that the impact of each additional iteration is relatively low. As
the computational time scales linearly with the number of iterations, we conclude
that convergence is achieved by as low as a single iteration.

\begin{figure}[ht!]
    \centering
    \includegraphics[width=\textwidth]{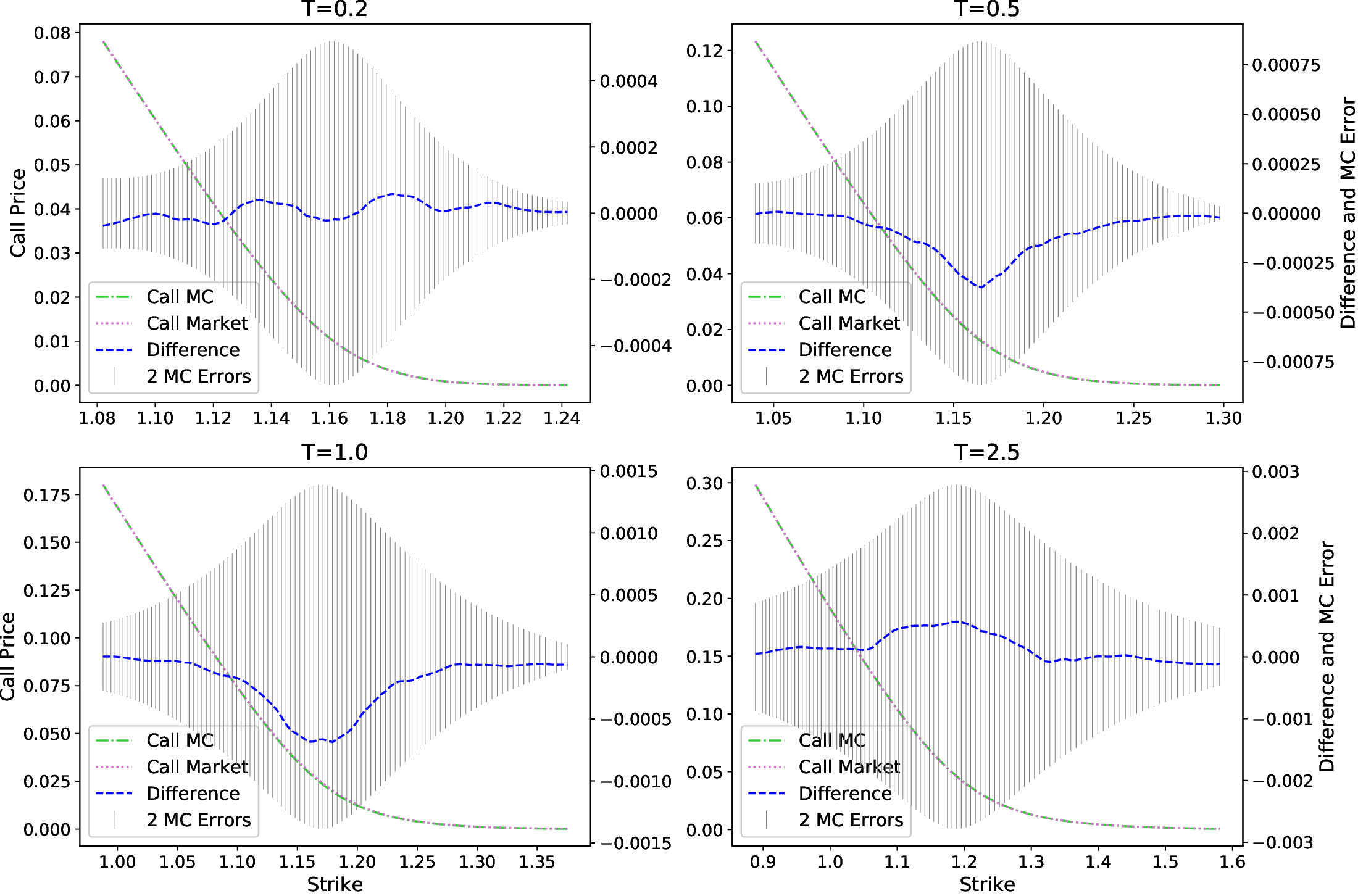}
    \caption{Comparing Monte Carlo prices to market prices of vanilla call
    options at several time slices. The differences are within two Monte Carlo
    error bounds.}
    \label{fig:callOptionsReprice}
\end{figure}

Finally, to validate the quality of the calibrated volatility surface, we
run a successive simulation of (\ref{eqn:LV2SR_LGM_forwardEuler}), with the same
setup we used for calibration, to price a range of vanilla call options with
expiries $T = 0.2, 0.5, 1.0, 2.5$. Figure \ref{fig:callOptionsReprice} shows
that the Monte Carlo prices match well with the market prices computed from the
input implied volatility surface. The differences seem to be within two Monte
Carlo errors.

\section{Discussion}
We derived the extension of the Dupire formula where the drift of the local
volatility model is given as a difference of two stochastic processes of general
form. The extended formula (\ref{eqn:extendeddupire}) can be used to calibrate
local volatility models with stochastic rates of this structure. We further
studied a general local volatility model where drift and diffusion are functions
of arbitrary number of stochastic factors, with the simple assumption that the
discount factor and the asset volatility are adapted functions of the It\^o
processes in the SDE system. The resulting generalized Dupire formula
(\ref{eqn:generaldupire}) can be used to calibrate a range of models, including
stochastic local volatility models with stochastic interest rates.
Both of these equations are given in implicit form, where the expectations on
the right hand side depend on the local volatility (leverage) function. Such
equations can be solved numerically by fixed-point iteration schemes. The expectations
appearing in these equations have no known analytical solutions, yet they can be
evaluated by numerical methods such as Monte Carlo or finite differences. We
presented a case study to demonstrate a calibration scheme by Monte Carlo
simulation. In the study we iteratively calibrated a local volatility model
subject to stochastic dometic and foreign interest rates, which were chosen to
follow an LGM process. The results show that the calibration is achieved after a
single iteration by a relatively low number of simulation paths, and that the
calibrated local volatility surface recovers market prices accurately.

\appendix
\section{Total implied variance surface formulation}

Quotes for various European call options with a range of strikes and maturities
are required for the evaluation of the Dupire formula (\ref{eqn:dupire}) or the
extended Dupire formula (\ref{eqn:extendeddupire}) to create a
\emph{local volatility surface}. In practice, one can create
a call price surface interpolator to evaluate the call price and its
derivatives in these equations along the grid where the local volatility surface
is being constructed. However the method for interpolation while evaluating the
Dupire formula is a concern as the interpolated values might introduce arbitrage
to the model. One way to address this problem is to construct a Black-Scholes
total implied variance surface and interpolate that instead.
As a matter of fact, practitioners typically work with market data that is in
the form of parametrized or dense implied volatility surfaces that are
calibrated with such penalty functions that aim to avoid or at least to minimize
arbitrage. The absence of
calendar spread arbitrage implies that the total implied variance surface is
a monotonically increasing function of time
\cite{Fengler2009,GatheralJacquier2014}. By construction, interpolating
the total implied variance surface and using these values in the Dupire
formula avoids calendar spread arbitrage. In this section we derive the
total implied variance parametrization of the extended Dupire formula.

The Black-Scholes European call option price function $C_{\text{BS}}$ can be
parametrized in terms of log-moneyness
\begin{equation*}
y(K, T) = \log\frac{K}{F_T},
\end{equation*}
where $F_T \equiv S_0 \frac{P^f(0, T)}{P^d(0, T)}$ is the forward price at time
$T$, and the total implied variance
\begin{equation*}
w(y(K, T), T) = \Sigma(K, T)^2 T
\end{equation*}
as \cite{Gatheral2012}
\begin{equation}
C_{\text{BS}}(P^d(0, T) F_T, y, w) = P^d(0, T) F_T \left(N(d_1) - e^y N(d_2)\right)\label{eqn:C_BS}
\end{equation}
with
\begin{equation*}
\begin{split}
d_1 =& -y w^{-\frac{1}{2}} + \frac{1}{2} w^{\frac{1}{2}},\\
d_2 =& d_1 - w^{\frac{1}{2}}.
\end{split}
\end{equation*}
Here $\Sigma(K, T)$ is the market implied volatility at strike $K$ and maturity
$T$, and $N(\cdot)$ is the standard Gaussian cumulative distribution function.
Noting that both $C_{\text{BS}}$ and $w$ depend on the strike $K$ indirectly
through $y(K, T)$, that is $C_{\text{BS}} = C_{\text{BS}}(P^d(0, T) F_T, y(K, T), w(y(K, T),
T))$, the first two derivatives of the call price with respect to the strike can
be computed as
\begin{equation*}
\begin{split}
\frac{\partial C_{\text{BS}}}{\partial K} =& \left(\frac{\partial C_{\text{BS}}}{\partial y}
+ \frac{\partial C_{\text{BS}}}{\partial w} \frac{\partial w}{\partial y} \right)
\frac{\partial y}{\partial K},\\
\frac{\partial^2 C_{\text{BS}}}{\partial K^2} =& \left[
\frac{\partial^2 C_{\text{BS}}}{\partial y^2}
+ \left(2\frac{\partial^2 C_{\text{BS}}}{\partial w \partial y} 
+ \frac{\partial^2 C_{\text{BS}}}{\partial w^2} \frac{\partial w}{\partial y}
\right) \frac{\partial w}{\partial y}
+ \frac{\partial C_{\text{BS}}}{\partial w} \frac{\partial^2 w}{\partial y^2}
\right] \left(\frac{\partial y}{\partial K}\right)^2\\
&+ \left(\frac{\partial C_{\text{BS}}}{\partial y}
+ \frac{\partial C_{\text{BS}}}{\partial w} \frac{\partial w}{\partial y} \right)
\frac{\partial^2 y}{\partial K^2}.
\end{split}
\end{equation*}
Since $\frac{\partial y}{\partial K} = \frac{1}{K}$ and
$\frac{\partial^2 y}{\partial K^2} = - \frac{1}{K^2}$ the second expression
can be written as
\begin{equation}
K^2 \frac{\partial^2 C_{\text{BS}}}{\partial K^2} =
\frac{\partial^2 C_{\text{BS}}}{\partial y^2}
+ \left(2\frac{\partial^2 C_{\text{BS}}}{\partial w \partial y} 
+ \frac{\partial^2 C_{\text{BS}}}{\partial w^2} \frac{\partial w}{\partial y}
- \frac{\partial C_{\text{BS}}}{\partial w}
\right) \frac{\partial w}{\partial y}
+ \frac{\partial C_{\text{BS}}}{\partial w} \frac{\partial^2 w}{\partial y^2}
-\frac{\partial C_{\text{BS}}}{\partial y}.\label{d2C_dK2}
\end{equation}
The right hand side of this equation demands evaluation of the derivatives of
the call price with respect to the log-moneyness and the total implied variance.
Using the identity $N'(d_1) = e^y N'(d_2)$ we compute the first $w$-derivative as
\begin{align}
\begin{split}
\frac{\partial C_{\text{BS}}}{\partial w}
=& P^d(0, T) F_T \left[N'(d_1) \frac{\partial d_1}{\partial w}
- e^y N'(d_2) \frac{\partial d_2}{\partial w}\right]\\
=&\frac{1}{2} P^d(0, T) F_T e^y N'(d_2) w^{-\frac{1}{2}};\label{eqn:dC_dw}
\end{split}\\
\intertext{and, since $N''(x) = - x N'(x)$, the second $w$-derivative
evaluates as}
\begin{split}
\frac{\partial^2 C_{\text{BS}}}{\partial w^2} =& \frac{1}{2} P^d(0, T) F_T e^y
\left[-N'(d_2) d_2 \frac{\partial d_2}{\partial w} w^{-\frac{1}{2}}
- \frac{1}{2} N'(d_2) w^{-\frac{3}{2}}\right]\\
=& \frac{1}{2}\frac{\partial C_{\text{BS}}}{\partial w}\left[-\frac{1}{4}
-\frac{1}{w} + \frac{y^2}{w^2}\right].\label{eqn:d2C_dw2}
\end{split}\\
\intertext{Furthermore, the remaining derivatives are}
\begin{split}
\frac{\partial^2 C_{\text{BS}}}{\partial w \partial y} =&
\frac{1}{2} P^d(0, T) F_T e^y N'(d_2) w^{-\frac{1}{2}}
\left[- d_2 \frac{\partial d_2}{\partial y} + 1\right]\\
=& \frac{\partial C_{\text{BS}}}{\partial w}
\left[-\frac{y}{w} + \frac{1}{2}\right],\label{eqn:d2C_dwdy}
\end{split}\\
\begin{split}
\frac{\partial C_{\text{BS}}}{\partial y} =&
P^d(0, T) F_T \left[N'(d_1) \frac{\partial d_1}{\partial y}
-e^y N(d_2) - e^y N'(d_2) \frac{\partial d_2}{\partial y} \right]\\
=& -P^d(0, T) F_T e^y N(d_2),\label{eqn:dC_dy}
\end{split}\\
\begin{split}
\frac{\partial^2 C_{\text{BS}}}{\partial y^2} =&
- P^d(0, T) F_T e^y \left[N(d_2) + N'(d_2) \frac{\partial d_2}{\partial y}\right]\\
=& \frac{\partial C_{\text{BS}}}{\partial y}
+ 2 \frac{\partial C_{\text{BS}}}{\partial w}.\label{eqn:d2C_dy2}
\end{split}
\end{align}
Plugging in equations (\ref{eqn:dC_dw}), (\ref{eqn:d2C_dw2}),
(\ref{eqn:d2C_dwdy}), (\ref{eqn:dC_dy}), and (\ref{eqn:d2C_dy2}) into
(\ref{d2C_dK2}) we arrive at
\begin{equation}
\frac{1}{2}K^2 \frac{\partial^2 C_{\text{BS}}}{\partial K^2} =
\frac{\partial C_{\text{BS}}}{\partial w} \left[
1 - \frac{y}{w} \frac{\partial w}{\partial y}
+ \frac{1}{2} \frac{\partial^2 w}{\partial y^2}
+ \frac{1}{4} \left(\frac{\partial w}{\partial y}\right)^2
\left(-\frac{1}{4}- \frac{1}{w} + \frac{y^2}{w^2}\right)
\right].\label{eqn:halfK2_d2C_dK2}
\end{equation}
Finally, we use the identities
\begin{equation*}
\begin{split}
\frac{\partial y}{\partial T}
=& -\frac{S_0}{F_T} \frac{\partial \frac{P^f(0, T)}{P^d(0, T)}}{\partial T}
= f^f(0, T) - f^d(0, T),\\
\frac{\partial (P^d(0, T) F_T)}{\partial T}
=& S_0 \frac{\partial (P^f(0, T))}{\partial T} = -f^f(0, T) P^d(0, T) F_T,
\end{split}
\end{equation*}
to formulate the time derivative of the call price as
\begin{equation}
\frac{\partial C_{\text{BS}}}{\partial T} = - f^f(0, T) C_{\text{BS}}
+ \frac{\partial C_{\text{BS}}}{\partial w} \frac{\partial w}{\partial T}
+\left(\frac{\partial C_{\text{BS}}}{\partial y}
+ \frac{\partial C_{\text{BS}}}{\partial w} \frac{\partial w}{\partial y}\right) (f^f(0, T)
- f^d(0, T)).\label{eqn:dC_dt}
\end{equation}
Plugging in equations (\ref{eqn:halfK2_d2C_dK2}) and (\ref{eqn:dC_dt}) into
(\ref{eqn:extendeddupire}) gives us the extended Dupire formula in the
log-moneyness/total implied variance parametrization.
\begin{equation}
\sigma^{\text{LV}}(K, T)^2 = \frac{
\frac{\partial C_{\text{BS}}}{\partial T}
- P^d(0, T) \mathbf{E}^{\mathbb{Q}^{\text{T}}}\left[(K r_T^d - S_T r_T^f) \mathds{1}_{S_T > K}\right]}
{\frac{\partial C_{\text{BS}}}{\partial w} \left[
1 - \frac{y}{w} \frac{\partial w}{\partial y} + \frac{1}{2} \frac{\partial^2 w}{\partial y^2}
+\frac{1}{4} \left(\frac{\partial w}{\partial y}\right)^2
\left(-\frac{1}{4}- \frac{1}{w} + \frac{y^2}{w^2}\right)
\right]},\label{eqn:extendeddupire_tiv}
\end{equation}
where the explicit forms of $C_{\text{BS}}$,
$\frac{\partial C_{\text{BS}}}{\partial w}$,
$\frac{\partial C_{\text{BS}}}{\partial y}$, and
$\frac{\partial C_{\text{BS}}}{\partial T}$ are
given by (\ref{eqn:C_BS}), (\ref{eqn:dC_dw}), (\ref{eqn:dC_dy}), and
(\ref{eqn:dC_dt}) respectively.

Equation (\ref{eqn:linkLV2SLV}) allows us to write the extended Dupire formula
for the two stochastic rates and stochastic local volatility model
(\ref{eqn:FXslvSDE}) in the total implied variance surface formulation as well.
Since the deterministic local volatility limiting case was already computed in
this formulation as in (\ref{eqn:extendeddupire_tiv}), we can write the leverage
function for the stochastic local volatility generalization as
\begin{equation}
L_s(K, T)^2 = \frac{
\frac{\partial C_{\text{BS}}}{\partial T}
- P^d(0, T) \mathbf{E}^{\mathbb{Q}^{\text{T}}}\left[(K r_T^d - S_T r_T^f) \mathds{1}_{S_T > K}\right]}
{\frac{\partial C_{\text{BS}}}{\partial w} \left[
1 - \frac{y}{w} \frac{\partial w}{\partial y} + \frac{1}{2} \frac{\partial^2 w}{\partial y^2}
+\frac{1}{4} \left(\frac{\partial w}{\partial y}\right)^2
\left(-\frac{1}{4}- \frac{1}{w} + \frac{y^2}{w^2}\right)
\right] \mathbf{E}^{\mathbb{Q}^{\text{T}}}\left[U_T | S_T=K\right]}.\nonumber
\end{equation}

\paragraph{Single stochastic rate limit}

In the limit where the foreign rates $r_T^f$ are deterministic, this equation becomes
\begin{equation}
\sigma^{\text{LV}}(K, T)^2 = \frac{
\frac{\partial C_{\text{BS}}}{\partial w} \frac{\partial w}{\partial T}
- f^d(0, T) \left(\frac{\partial C_{\text{BS}}}{\partial y}
+ \frac{\partial C_{\text{BS}}}{\partial w} \frac{\partial w}{\partial y}\right)
- P^d(0, T) K \mathbf{E}^{\mathbb{Q}^{\text{T}}}\left[ r_T^d \mathds{1}_{S_T > K}\right]}
{\frac{\partial C_{\text{BS}}}{\partial w} \left[
1 - \frac{y}{w} \frac{\partial w}{\partial y} + \frac{1}{2} \frac{\partial^2 w}{\partial y^2}
+\frac{1}{4} \left(\frac{\partial w}{\partial y}\right)^2
\left(-\frac{1}{4}- \frac{1}{w} + \frac{y^2}{w^2}\right)
\right]}. \label{eqn:dupire_single_tiv}
\end{equation}

\paragraph{Deterministic rates limit}

In the limit where both the domestic rates $r_T^d$ and the foreign rates $r_T^f$
are deterministic, the equation further simplifies to the form given in
\cite{Gatheral2012}
\begin{equation}
\sigma^{\text{LV}}(K, T)^2 = \frac{\frac{\partial w}{\partial T}}
{1 - \frac{y}{w} \frac{\partial w}{\partial y}
+ \frac{1}{2} \frac{\partial^2 w}{\partial y^2}
+ \frac{1}{4} \left(\frac{\partial w}{\partial y}\right)^2
\left(-\frac{1}{4}- \frac{1}{w} + \frac{y^2}{w^2}\right)
}. \label{eqn:dupire_det_tiv}
\end{equation}

\section*{Acknowledgments}
The authors are grateful to the SIAM SIFIN reviewers and the editor for their valuable
comments that have substantially improved the paper. The authors are indebted to
Dooheon Lee for numerous enlightening discussions and guidance about the
construction and the flow of this paper.
The authors would also like to thank Agus Sudjianto for supporting
this research, and Vijayan Nair for suggestions, feedback, and discussion
regarding this work. Orcan Ogetbil is appreciative for Paul Feehan's original
introduction to the subject and his endorsement. Any opinions, findings and
conclusions or recommendations expressed in this material are those of the
author and do not necessarily reflect the views of Wells Fargo Bank, N.A., its
parent company, affiliates and subsidiaries.

\bibliographystyle{unsrt}
\bibliography{localvol}

\end{document}